\documentclass[conference]{IEEEtran}

\usepackage{arydshln}

\usepackage[noadjust]{cite}
\usepackage{enumitem}

\usepackage{xcolor}

\definecolor{ieeeblue}{RGB}{0,120,177}
\usepackage[colorlinks = true,
            linkcolor= ieeeblue,
            urlcolor  = ieeeblue,
            citecolor = ieeeblue,
            anchorcolor = ieeeblue]{hyperref}
\usepackage{nameref}
\usepackage{amsmath,amssymb,amsfonts}
\usepackage{algorithmic}
\usepackage{graphicx}
\graphicspath{{./images/}}
\usepackage{textcomp}
\usepackage[framemethod=TikZ]{mdframed}   % for framing
\usepackage{float}
\usepackage{babel}
\usepackage{array,tabularx,multirow,makecell,booktabs,flafter}
\newcolumntype{M}[1]{>{\centering\arraybackslash}m{#1}}
\usepackage{setspace}
\usepackage{balance}

\usepackage{varwidth}
\usepackage{verbatim}
\usepackage{standalone}
\usepackage{tikz}
\usetikzlibrary{positioning,shapes.geometric,fit,calc}
\usepackage{pgfplots}
\pgfplotsset{compat=1.16,  every non boxed x axis/.append style={x axis line style=-},
     every non boxed y axis/.append style={y axis line style=-}}

\usepackage{pifont}

\usepackage{xspace}
\newcommand{\etal}{\textit{et al.}~}

\newcommand{\one}{({\textit{i}})\xspace}
\newcommand{\two}{({\textit{ii}})\xspace}
\newcommand{\three}{({\textit{iii}})\xspace}
\newcommand{\four}{({\textit{iv}})\xspace}
\newcommand{\five}{({\textit{v}})\xspace}

\makeatletter
\renewcommand{\paragraph}[1]{\vspace*{0.03in}\noindent{\bf #1.}\hspace{0.25ex \@plus1ex \@minus.2ex}}
\makeatother

\definecolor{grey}{cmyk}{0,0,0,.1}
\definecolor{black}{cmyk}{0,0,0,1}

\newenvironment{highlightBox}[1][]{
    \ifstrempty{#1}{
        \mdfsetup{
            nobreak=true,
            frametitle={
                \tikz[baseline=(current bounding box.east), outer sep=0pt]
                \node[anchor=east, rectangle, fill=black] {\strut \sffamily \textcolor{white}{Titel}};
            }
        }
    }{
        \mdfsetup{
            frametitle={
                \tikz[baseline=(current bounding box.east), outer sep=0pt]
                \node[anchor=east, rectangle, fill=black] {\strut \sffamily \textcolor{white}{#1}};
            }
        }
    }

    \mdfsetup{
        innertopmargin=10pt,
        linecolor=black,
        linewidth=2pt,
        topline=true,
        innertopmargin=8,
        innerbottommargin=8,
        innerleftmargin=8,
        innerrightmargin=8,
        skipabove=12,
        skipbelow=12,
        frametitleaboveskip=\dimexpr-\ht\strutbox\relax
    }

    \begin{mdframed}[]
}{
    \end{mdframed}
}

\usepackage{fancyhdr}
\begin{document}
\title{SoK: Bridging Trust into the Blockchain \\A~Systematic Review on On-Chain Identity}

\author{\IEEEauthorblockN{Awid Vaziry}
\IEEEauthorblockA{\textit{Service-centric Networking}\\
\textit{Technische Universität Berlin}\\
Berlin, Germany\\
vaziry@tu-berlin.de}
\and
\IEEEauthorblockN{Kaustabh Barman}
\IEEEauthorblockA{\textit{Service-centric Networking}\\
\textit{Technische Universität Berlin}\\
Berlin, Germany\\
kaustabh.barman@tu-berlin.de}
\and
\IEEEauthorblockN{Patrick Herbke}
\IEEEauthorblockA{\textit{Service-centric Networking}\\
\textit{Technische Universität Berlin}\\
Berlin, Germany\\
p.herbke@tu-berlin.de}}

\maketitle

%% ABSTRACT
% As a general rule, do not put math, special symbols or citations in the abstract or keywords.

\begin{abstract}
% Context 
The ongoing regulation of blockchain-based services and applications requires the identification of users who are issuing transactions on the blockchain.
% Objectives 
This systematic review explores the current status, identifies research gaps, and outlines future research directions for establishing trusted and privacy-compliant identities on the blockchain (on-chain identity).
% Methods 
A systematic search term was applied across various scientific databases, resulting in the collection of 2232 potentially relevant research papers. These papers were narrowed down in two methodologically executed steps to 98 and finally to 13 relevant sources. The relevant articles were then systematically analyzed based on a set of screening questions.
% Results 
The results of the selected studies have provided insightful findings on the mechanisms of on-chain identities. On-chain identities are established using zero-knowledge proofs, public key infrastructure/certificates, and web of trust approaches. The technologies and architectures used by the authors are also highlighted. Trust has emerged as a key research gap, manifesting in two ways: firstly, a gap in how to trust the digital identity representation of a physical human; secondly, a gap in how to trust identity providers that issue identity confirmations on-chain.
% Conclusions 
Potential future research avenues are suggested to help fill the current gaps in establishing trust and on-chain identities.
\end{abstract}


\IEEEpeerreviewmaketitle

\section{Introduction}
\label{sec:introduction}

\IEEEPARstart{W}{ith} the increasing adoption and institutionalization of blockchain applications and decentralized finance services, regulators require proof of user identities to prevent fraud and maintain the integrity of financial markets. Beyond privacy and data protection concerns, there is the broader challenge of bridging identity representations and making them available on-chain \cite{wronka_financial_2023, liu_blockchain-based_2020}.

Blockchain transactions and consumer-facing decentralized finance applications have gained interest in recent years. However, current blockchain applications often do not comply with existing financial regulations, such as anti-money laundering (AML), know-your-customer (KYC), and counter-terrorist financing (CTF) rules. Alignment with these regulatory frameworks requires blockchain-based verification of user identities. Such alignment is crucial to prevent potential bans on blockchain applications and foster broader adoption and innovation \cite{wronka_financial_2023,2024_tomonori_regulation}. This study aims to work towards the development of a trusted and usable on-chain identity solution. These on-chain identity solutions typically combine and integrate a variety of technologies, resulting in a hybrid of on-chain (blockchain-based) and off-chain (internet- and physical-world-based) components \cite{lesavre_taxonomic_nodate}.

In this paper we conduct a structured systematic review using the guidelines for systematic reviews in software engineering as proposed by Kitchenham~\etal \cite{kitchenham_guidelines_2007} and the \textit{Preferred Reporting Items for Systematic Reviews and Meta-Analyse} \cite{page_prisma_2021} (PRISMA) framework. The objective is to synthesize and critically evaluate the existing body of knowledge on blockchain-based on-chain identity solutions. This involves aggregating the technologies and architectures in use and identifying common patterns and shortcomings. Based on this analysis, we identify research gaps and suggest future avenues of investigation. Although there are existing reviews in the general context of using blockchain for identity management, this is, to the best of our knowledge, the first structured literature review with a specific focus on on-chain identities.

\section{Background}
\label{sec:background}
This section introduces the preliminary concepts essential for comprehending the review and its outcomes. Consequently, blockchain, identity, and related technologies, like public key infrastructure and zero-knowledge proofs, are presented. Furthermore, other surveys are introduced, and this work is situated within the context of existing literature reviews on blockchain and identity.

\subsection{Blockchain}
A blockchain is a specific instance of a distributed ledger, a technological concept that enables the maintenance of append-only transactional databases in a decentralized manner. Bitcoin first introduced securing a virtual currency without a middleman and a central point of failure~\cite{nakamoto2008bitcoin}. Nodes in the network bundle transactions into a block and agree on including a block by adhering to a consensus algorithm. In 2015, \textit{Buterin}~\etal introduced Ethereum, which offers Turing complete code execution, so-called smart contracts~\cite{buterin2014next}. Blockchains can be classified into two categories regarding governance in the network - permissionless and permissioned networks. In \textit{permissionless blockchains}, anyone can run a node and secure the network. In most cases, honest behaving nodes are rewarded by economic incentives; dishonest behavior is usually punished or economically disadvantageous. Fault tolerance mechanisms ensure that a minority of nodes do not possess sufficient power to impair execution. To prevent spam and to reward node operators for computation and storage, users are typically required to pay to use the blockchain. In Ethereum, payment is denominated in a unit called gas, which limits how much can be spent in a transaction. This leads to resource constraints and limited throughput. \textit{Permissioned blockchains} are secured by a single entity or a consortium, and access to the network is usually restricted. Hyperledger-based distributed ledger technologies like Fabric and Aries are examples of permissioned blockchains \cite{palma_what_2021}. 

This work will refer to both on-chain and off-chain approaches. In this discussion, we define these two terms as follows:
\begin{itemize}
    \item \textit{On-Chain: }refers to operations carried out or recorded on the blockchain~\cite{eberhardt2017or}. Data are stored in the nodes of the distributed ledger network and can be read by anyone and used by on-chain smart contracts. 
    \item \textit{Off-Chain: }describes any data, application, or computations that occur outside of the blockchain~\cite{eberhardt2017or}. For example on a web server or a local machine.
\end{itemize}

\subsection{Identity and Self Sovereign Identity}

The International Organization for Standardization defines identity as a "set of attributes related to an entity." Identity attributes are characteristic properties of an entity, such as a device's personal ID number, address, or MAC address. An identity can be represented by a credential (e.g., digital record, password, physical ID card)~\cite{iso_24760-1:2019}.

The shift to web3, with its decentralized and user-centric applications, has pushed the adoption of Self-Sovereign Identity (SSI) systems. SSI enables individuals to control their identities without central authorities, for instance, using blockchain and distributed ledger technologies for secure, tamper-proof identity verification~\cite{dunphy2018first}.

The issuer-holder-verifier framework is commonly used for SSI and blockchain-based identity management and verification. The issuer creates and issues credentials to the holder. The holder controls the credentials and determines when and with whom to share them while maintaining privacy and autonomy. The verifier validates the credentials, ensuring their legitimacy without requiring direct interaction with the issuer.~\cite{muhle2018survey}.

\subsection{Public Key Infrastructure}
Public Key Infrastructure (PKI) is a framework designed to manage digital keys and certificates, ensuring electronic communications security and authenticity. PKI is built upon several key components, such as trusted entities named Certificate Authorities (CAs), Registration Authorities (RAs), Digital Certificates, and Certificate Revocation Lists (CRLs). A certificate is a document that binds public keys to the identities of individuals or entities. X.509 certificates are the defined standard format for public key certificates within the PKI, containing the public key of the certificate owner, identifying information, and the digital signature of the issuing CA~\cite{cooper2008internet}.

\subsection{Zero-Knowledge Proofs}

A zero-knowledge proof (zkp) is a cryptographic technique that enables one party (the prover) to prove to another party (the verifier) that a statement is true without revealing any additional information. Some basic principles of Zero-Knowledge Proof as laid down by \textit{Goldwasser}~\etal~\cite{goldwasser2019knowledge} are as follows: 
\begin{itemize}
    \item \textit{Completeness:} A truthful prover can convince the verifier if the statement is true.
    \item \textit{Soundness:} If the statement is false, no cheating prover can convince the verifier of its truth.
    \item \textit{Zero-Knowledge:} If the statement is true, the verifier learns nothing other than the statement is true.
\end{itemize}
A zk-SNARK (zero-knowledge succinct non-interactive arguments of knowledge) is a zkp with the additional properties of non-interactivity and succinctness. A non-interactive zero-knowledge proof (NIZK) eliminates the need for interaction between the prover and the verifier by generating a concise proof that the verifier can verify independently. The concept of succinctness implies that proofs are smaller in size than the propositions they prove \cite{ben-sasson_succinct_2014}.
tions, enabling trusted interactions over untrusted networks.

\subsection{Related Work}
Related literature reviews on the intersection of blockchain and identity can be divided into three categories. First, specialized surveys focusing on specific areas, such as blockchain and identity for healthcare \cite{houtan_survey_2020} or the Internet of Things \cite{huo_comprehensive_2022}. Second, general surveys aimed at collecting all available literature. \textit{Liu}~\etal~\cite{liu_blockchain-based_2020} and \textit{Ahmed}~\etal~\cite{9927415} provide such a general survey in 2020 and 2023, respectively. \textit{Ahmed}~\etal concludes that blockchain technology can significantly improve identity management systems by enhancing security, privacy, and user control. However, it also notes that integration is still in its early stages. Our work is unique due to the focus on on-chain verifiable identities, i.e., identities made for use on the blockchain. To the best of our knowledge and belief, this is the first review of its kind.

\subsection{Definition of on-chain identity}
This work focuses on on-chain identity. The term is not formally defined in academia and was first used by \textit{Azouvi}~\etal~\cite{azouvi_who_2017} in 2017 within a context that aligns with the scope of our work.
To comply with data protection regulations (e.g. GDPR), our definition of on-chain identity explicitly excludes the storage of personal data on the blockchain, even in encrypted form.
We define on-chain identity as the possession of an attestation (e.g. a token or claim) by an externally owned account (a user), which can be presented to an on-chain entity (a smart contract), to demonstrate that the users identity has been verified. In simple terms, proofing the possession of a verified identity on the blockchain.
\section{Method of literature review}
\label{sec:method}
This review employs a systematic approach to retrieve, filter, and select relevant publications. This approach is primarily based on the updated guidelines for Preferred Reporting Items for Systematic Reviews and Meta-Analyses (PRISMA) methodology by \textit{Page}~\etal~\cite{page_prisma_2021}. PRISMA is renowned for its efficiency in enhancing the quality and transparency of review reporting. To refine the method, the eight-step guide for information systems research, as proposed by \textit{Okoli}~\etal~\cite{okoli_guide_2015} is integrated as well as the guidelines for systematic reviews in software engineering as proposed by \textit{Kitchenham}~\etal~\cite{kitchenham_guidelines_2007}. These guidelines address the planning, conducting, and reporting phases of systematic literature reviews (SLRs). The guidelines address the development of a review protocol, the identification and selection of studies, the assessment of study quality, the extraction and synthesis of data, and the reporting of findings. Both guidelines highlight the importance of pre-planning the review, primarily through creating a thorough review protocol. The protocol ensures the accuracy and significance of the results by minimizing ad-hoc decisions.

The review methodology is structured to provide a comprehensive understanding of the procedure. The subsequent chapters are structured in accordance with the PRISMA framework, which details the \one search process, \two selection processes, \three eligibility check, and \four screening procedures. The systematic methodology will be critically evaluated, and the findings will be presented at a high level. A thorough analysis and discussion of these findings will be undertaken in the subsequent chapter \ref{sec:discussion}.

\subsection{Objective definition and planning of the Review}
The primary objective of this systematic literature review is to establish the current state of the art of on-chain verifiable identities. Therefore, this work systematically collects, evaluates, and synthesizes the findings from existing research on blockchain-based identity solutions. A specific focus is placed on the technologies used and the identification and classification of components for establishing identities, emphasizing those implemented on-chain and off-chain. Three research questions, which serve as a framework for the review, are developed based on the current state of the art and questions deemed relevant for the review.
Currently known blockchain identity protocols have at least some of their logic being computed or stored off-chain. The first research question [\ref{RQ1}] explores the current state of knowledge regarding the realization of on-chain identities concerning implementing on-chain logic. The second research question [\ref{RQ2}] further investigates the additional algorithms or technologies used. This research question aims to investigate whether there are patterns of approaches deemed useful by different researchers. The final research question [\ref{RQ3}] explores establishing trust in the system. Since the blockchain is a deterministic state machine that operates on data already present on the blockchain or new data published during transactions, initiating trust is challenging. The question arises: Was a new identity submitted on the blockchain approved by the real-world owner of this identity?

\begin{enumerate}[before=\vspace{10pt}, after=\vspace{10pt}, itemsep=12pt, label=\textbf{RQ\arabic*}]
    \item  \label{RQ1} Which components of blockchain-based identity solutions are currently implemented on-chain versus off-chain, and what are the underlying reasons for their respective implementations?
    
    \item \label{RQ2} What additional technologies, algorithms, and patterns are frequently used for on-chain and off-chain components of blockchain-based identity solutions, and what criteria or considerations guide their selection?
    
    \item \label{RQ3} How is trust established within a solution, specifically addressing credential issuance and the integration of real-world data/identities into on-chain environments?
\end{enumerate}

\subsection{Step 1 - Search Process}

The initial step in the PRISMA process is the identification of pertinent literature. The primary search terms are \textit{`blockchain'} and \textit{`identity management'}. These are expanded by including synonyms, abbreviations, and domain-specific concepts. The case-insensitive search term employed is as follows:

\begin{hyphenrules}{nohyphenation}

\begin{highlightBox}[Search Term]
  \begin{minipage}{\linewidth} % Verwenden Sie \linewidth, um die Breite der minipage auf die Breite der Box einzustellen.
    \texttt{("\textbf{blockchain}" OR "\textbf{distributed~ledger}" OR "\textbf{on-chain}" OR "\textbf{decentralized~finance}" OR "\textbf{defi}" OR  "\textbf{smart~contract}") 
    \\ AND \\
    ("\textbf{identity management}" OR "\textbf{verifiable credentials}" OR "\textbf{self-sovereign identity}" OR "\textbf{SSI}" OR "\textbf{decentralized~identifiers}" OR "\textbf{know~your~customer}"  OR "\textbf{KYC}")
}
  \end{minipage}
\end{highlightBox}

\end{hyphenrules}

The researchers conducted a literature search across six databases recognized as comprehensive computer science and technical research repositories. One of those databases (arXiv) contains grey literature, which is acceptable by our guidelines and due to the multi-step filtering and quality check process. The search was executed on February 26, 2024. When the option to filter the search was available, the search was conducted on titles and abstracts, thereby enhancing the relevance of retrieved documents. The search was limited to literature published in 2017 and subsequent years, which aligns with significant advancements in the blockchain and Ethereum ecosystem \cite{schar_decentralized_2020}. The combined use of all databases yielded a total of \textbf{2,232} publications. For the most recent outcomes of the query, please consult the following hyperlinks:

\begin{itemize}
\item IEEE Xplore: \href{https://ieeexplore.ieee.org/search/searchresult.jsp?action=search&matchBoolean=true&queryText=(%22Abstract%22:%22blockchain%22%20OR%20%22Abstract%22:%22distributed%20ledger%22%20OR%20%22Abstract%22:%22on-chain%22%20OR%20%22Abstract%22:%22decentralized%20finance%22%20OR%20%22Abstract%22:%22defi%22%20OR%20%22Abstract%22:%22smart%20contract%22)%20AND%20(%22Abstract%22:%22self-sovereign%20identity%22%20OR%20%22Abstract%22:%22SSI%22%20OR%20%22Abstract%22:%22decentralized%20identifiers%22%20OR%20%22Abstract%22:%22KYC%22%20OR%20%22Abstract%22:%22know%20your%20customer%22%20OR%20%22Abstract%22:%22verifiable%20credentials%22%20OR%20%22Abstract%22:%22Identity%20Management%22)%20OR%20(%22Document%20Title%22:%22blockchain%22%20OR%20%22Document%20Title%22:%22distributed%20ledger%22%20OR%20%22Document%20Title%22:%22on-chain%22%20OROR%20%22Document%20Title%22:%22decentralized%20finance%22%20OR%20%22Document%20Title%22:%22defi%22%20OR%20%22Document%20Title%22:%22smart%20contract%22)%20AND%20(%22Document%20Title%22:%22self-sovereign%20identity%22%20OR%20%22Document%20Title%22:%22SSI%22%20OR%20%22Document%20Title%22:%22decentralized%20identifiers%22%20OR%20%22Document%20Title%22:%22KYC%22%20OR%20%22Document%20Title%22:%22know%20your%20customer%22%20OR%20%22Document%20Title%22:%22verifiable%20credentials%22%20OR%20%22Document%20Title%22:%22Identity%20Management%22)&highlight=true&returnFacets=ALL&returnType=SEARCH&matchPubs=true&ranges=2017_2024_Year}{IEEE-1}

\item SpringerLink:
\href{https://link.springer.com/search?query=("blockchain"+OR+"distributed+ledger"+OR+"on-chain"+OR+"decentralized+finance"+OR+"defi"+OR+"smart+contract")+AND+("self-sovereign+identity"+OR+"SSI"+OR+"decentralized+identifiers"+OR+"KYC"+OR+"know+your+customer"+OR+"verifiable+credentials"+OR+"Identity+Management")&facet-language="En"&facet-content-type="ConferencePaper"&facet-end-year=2024&showAll=true&date-facet-mode=between&facet-start-year=2017}{Springer-1}
\href{https://link.springer.com/search?query=("blockchain"+OR+"distributed+ledger"+OR+"on-chain"+OR+"decentralized+finance"+OR+"defi"+OR+"smart+contract")+AND+("self-sovereign+identity"+OR+"SSI"+OR+"decentralized+identifiers"+OR+"KYC"+OR+"know+your+customer"+OR+"verifiable+credentials"+OR+"Identity+Management")&facet-language="En"&facet-end-year=2024&showAll=true&date-facet-mode=between&facet-start-year=2017&facet-content-type="Article"}{Springer-2}

\item ACM Digital Library:
\href{https://dl.acm.org/action/doSearch?fillQuickSearch=false&target=advanced&expand=dl&AllField=Abstract%3A(("blockchain"+OR+"distributed+ledger"+OR+"on-chain"+OR+"decentralized+finance"+OR+"defi"+OR+"smart+contract")+AND+("self-sovereign+identity"+OR+"SSI"+OR+"decentralized+identifiers"+OR+"KYC"+OR+"know+your+customer"+OR+"verifiable+credentials"+OR+"Identity+Management"))}{ACM-1}
\href{https://dl.acm.org/action/doSearch?fillQuickSearch=false&target=advanced&expand=dl&AllField=Title%3A(("blockchain"+OR+"distributed+ledger"+OR+"on-chain"+OR+"decentralized+finance"+OR+"defi"+OR+"smart+contract")+AND+("self-sovereign+identity"+OR+"SSI"+OR+"decentralized+identifiers"+OR+"KYC"+OR+"know+your+customer"+OR+"verifiable+credentials"+OR+"Identity+Management"))+AND+Title%3A(("blockchain"+OR+"distributed+ledger"+OR+"on-chain"+OR+"decentralized+finance"+OR+"defi"+OR+"smart+contract")+AND+("self-sovereign+identity"+OR+"SSI"+OR+"decentralized+identifiers"+OR+"KYC"+OR+"know+your+customer"+OR+"verifiable+credentials"+OR+"Identity+Management"))+AND+Title%3A(("blockchain"+OR+"distributed+ledger"+OR+"on-chain"+OR+"decentralized+finance"+OR+"defi"+OR+"smart+contract")+AND+("self-sovereign+identity"+OR+"SSI"+OR+"decentralized+identifiers"+OR+"KYC"+OR+"know+your+customer"+OR+"verifiable+credentials"+OR+"Identity+Management"))+AND+Title%3A(("blockchain"+OR+"distributed+ledger"+OR+"on-chain"+OR+"decentralized+finance"+OR+"defi"+OR+"smart+contract")+AND+("self-sovereign+identity"+OR+"SSI"+OR+"decentralized+identifiers"+OR+"KYC"+OR+"know+your+customer"+OR+"verifiable+credentials"+OR+"Identity+Management"))}{ACM-2}

\item Wiley Online Library:
\href{https://onlinelibrary.wiley.com/action/doSearch?AfterMonth=&AfterYear=2017&BeforeMonth=3&BeforeYear=2024&Ppub=&field1=Abstract&field2=AllField&field3=AllField&text1=("blockchain"+OR+"distributed+ledger"+OR+"on-chain"+OR+"decentralized+finance"+OR+"defi"+OR+"smart+contract")+AND+("self-sovereign+identity"+OR+"SSI"+OR+"decentralized+identifiers"+OR+"know+your+customer"+OR+"KYC"++OR+"verifiable+credentials"+OR+"Identity+Management")&text2=&text3=&startPage=&PubType=journal}{Wiley-1}

\item ScienceDirect: 
\href{https://www.sciencedirect.com/search?date=2017-2024&tak=("blockchain" OR "distributed ledger" OR "on-chain" OR "decentralized finance" OR "defi" OR "smart contract") AND ("self-sovereign identity" OR "SSI")}{SD-1}
\href{https://www.sciencedirect.com/search?date=2017-2024&tak=("blockchain" OR "distributed ledger" OR "on-chain" OR "decentralized finance" OR "defi" OR "smart contract") AND ("KYC" OR "know your customer")}{SD-2}
\href{https://www.sciencedirect.com/search?date=2017-2024&tak=("blockchain" OR "distributed ledger" OR "on-chain" OR "decentralized finance" OR "defi" OR "smart contract") AND ("decentralized identifiers")}{SD-3}
\href{https://www.sciencedirect.com/search?date=2017-2024&tak=("blockchain" OR "distributed ledger" OR "on-chain" OR "decentralized finance" OR "defi" OR "smart contract") AND ("verifiable credentials" OR "Identity Management")&years=2024&lastSelectedFacet=years}{SD-4}

\item arXiv e-Print archive:
\href{https://arxiv.org/search/advanced?advanced=&terms-0-operator=AND&terms-0-term="blockchain"+OR+"distributed+ledger"+OR+"on-chain"+OR+"decentralized+finance"+OR+"defi"+OR+"smart+contract"&terms-0-field=title&terms-1-operator=AND&terms-1-term="self-sovereign+identity"+OR+"SSI"+OR+"decentralized+identifiers"+OR+"know+your+customer"+OR+"KYC"++OR+"verifiable+credentials"+OR+"Identity+Management"&terms-1-field=title&classification-physics_archives=all&classification-include_cross_list=include&date-filter_by=all_dates&date-year=&date-from_date=&date-to_date=&date-date_type=submitted_date&abstracts=show&size=50&order=-announced_date_first}{arXiv-1}
\href{https://arxiv.org/search/advanced?advanced=&terms-0-operator=AND&terms-0-term="blockchain"+OR+"distributed+ledger"+OR+"on-chain"+OR+"decentralized+finance"+OR+"defi"+OR+"smart+contract"&terms-0-field=abstract&terms-1-operator=AND&terms-1-term="self-sovereign+identity"+OR+"SSI"+OR+"decentralized+identifiers"+OR+"know+your+customer"+OR+"KYC"++OR+"verifiable+credentials"+OR+"Identity+Management"&terms-1-field=abstract&classification-physics_archives=all&classification-include_cross_list=include&date-filter_by=all_dates&date-year=&date-from_date=&date-to_date=&date-date_type=submitted_date&abstracts=show&size=100&order=-announced_date_first}{arXiv-2}
\end{itemize}

The search results from the \textit{IEEE Xplore} and \textit{ACM Digital Library} were exported as a spreadsheet (\texttt{.csv} file). The outcomes of the \textit{SpringerLink} and \textit{ScienceDirect} databases were exported as a BibTeX library (\texttt{.bib} file). The \textit{ArXiv} and \textit{Wiley} results were added manually. All results were processed with a data analysis and manipulation library called \texttt{pandas}. The data was cleaned and aggregated, and \textbf{31} duplicates were removed.

The results of our searches were exported into a spreadsheet and subsequently enriched by 
\one document title,
\two abstract,
\three author names,
\four publication year, and
\five Digital Object Identifier (DOI).

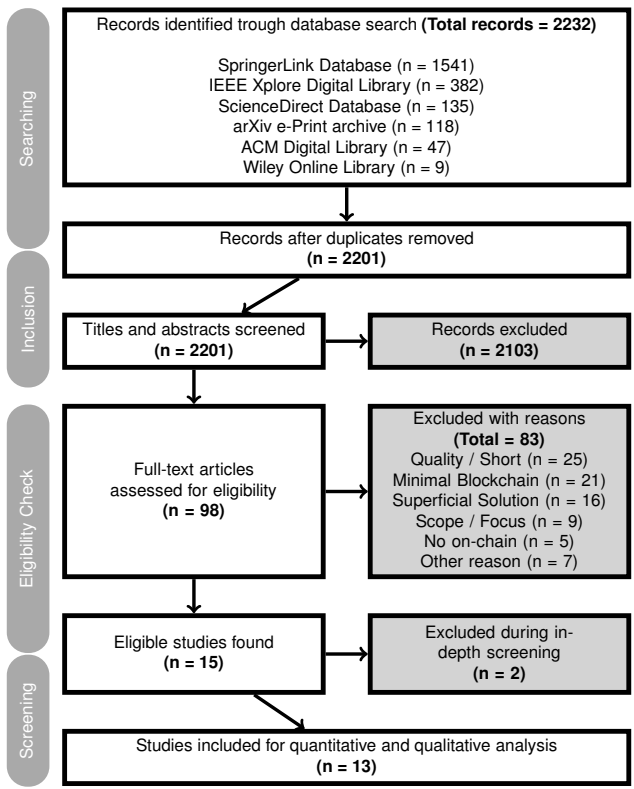
\begin{figure}[!ht]
    \centering
    \resizebox{\columnwidth}{!}{
        % colors
\definecolor{lightgray}{RGB}{211,211,211} %light gray for contrast
\definecolor{darkgray}{RGB}{169,169,169} %dark gray for visibility
\definecolor{gray}{RGB}{128,128,128} %normal gray
\definecolor{black}{RGB}{0,0,0} %black

\newcommand\x{width{"Records identified trough database search  Total records   2232 "}}

\begin{tikzpicture}[thick, node distance=2em]

%style for all rec
\tikzset{
  rec/.style={
    rectangle,
    fill=white, % Fill with white for high contrast
    draw=black, % Draw border with black color
    line width=1mm, % Adjust border width
    text=black, % Text color set to black for readability
    /utils/exec={\sffamily},
    minimum width=\x*1.2
  }
}

%style for small rec 
\tikzset{
  srec/.style={
    minimum width=\x*0.55
  }
}

%style for small rec with adapted width
\tikzset{
  svrec/.style={
    minimum width=\x*0.55,
    text width = \x*0.5,
    align=center
  }
}

%style for excluded
\tikzset{
  excl/.style={
    fill=lightgray, % Light gray fill for excluded elements
    text=black, % Text color set to black for readability
  }
}

\tikzset{
    barstyle/.style={
        fill=darkgray, % Dark gray fill for bar elements
        text=white, % Text color set to white for readability
        /utils/exec={\sffamily},
        align=center,
        inner ysep=6pt,
        rotate=90,
        rounded corners=1.2em,
        minimum height=2.5em
    }
}

%1DB
\node[rec]
  (DB) {\begin{tabular}{c}Records identified trough database search \textbf{(Total records = 2232)} \\ \\ SpringerLink Database (n = 1541) \\ IEEE Xplore Digital Library (n = 382) \\ ScienceDirect Database (n = 135) \\ arXiv e-Print archive (n = 118) \\  ACM Digital Library (n = 47) \\  Wiley Online Library (n = 9) \end{tabular}};

%2 dup
\node[rec, 
    below= of DB.south]
  (dup) {\begin{tabular}{c} Records after duplicates removed \\ \textbf{(n = 2201)} \end{tabular}};
  
%1+2 label: ID
\path let
\p1=(dup), \p2=(DB.north)
in
node[barstyle,left=of DB.north west,minimum width=\y2-\y1-0.1em] 
  (barID) {Searching};

%3 screen+ex  
\path (dup.north) -- (dup) coordinate[midway] (scex);

\node[rec, srec, below= of dup.south west,anchor=north west]
  (screen) {\begin{tabular}{c} Titles and abstracts screened \\ \textbf{(n = 2201)} \end{tabular}};
  
\node[rec, srec, excl, below= of dup.south east,anchor=north east]
  (ex) {\begin{tabular}{c} Records excluded \\ \textbf{(n = 2103)} \end{tabular}};

%2+3 label-Screening: screenInfo
%\begin{comment}
\path let
\p1=(screen.south), \p2=(dup)
in
node[barstyle,left=of dup,minimum width=\y2-\y1+1em] 
  (barScreen) {Inclusion};
%\end{comment}

%4 FTin+FTout
\path (screen.south) -- (ex.south) coordinate[midway] (scex);  

%height muss angepasst werden
\node[rec, excl, svrec, below=of ex, minimum height=10em]
  (FTout) {\mbox{Excluded with~reasons} \textbf{(Total = 83)}  \\ Quality / Short
  (n = 25) \\ Minimal Blockchain (n = 21) \\ Superficial Solution (n = 16) \\ Scope / Focus (n = 9) \\  No on-chain (n = 5) \\ Other reason (n = 7) };
%Minimal Blockchain: 21, Quality: 20 + 5 short, superficial: 16, scope: 6 + focus:3, no on-chain: 5, other: 7, . Später noch 1 no on-chain und 1 scope exclude

%height muss angepasst werden
\node[rec, svrec, below=of screen, minimum height=10em]
  (FTin) { Full-text articles assessed for eligibility \\ \textbf{(n = 98)}};

%5 Elli+Elliext  %height muss angepasst werden
\node[rec, srec, below=of FTin, minimum height=4.5em]
  (Elli) {\begin{tabular}{c} Eligible studies found \\ \textbf{(n = 15)} \end{tabular}};
  
\node[rec, svrec, excl, below=of FTout, minimum height=4.5em]
  (Elliext) { Excluded during in-depth screening\\ \textbf{(n = 2)}};

%4+5 label: Eligibility
\path let
\p1=(Elli), \p2=(FTin.north)
in
node[barstyle,left=of FTin.north west,minimum width=\y2-\y1-0.1em] 
  (barElli) {Eligibility Check};

%6 final
\path (Elli.south) -- (Elliext.south) coordinate[midway] (FT);
\node[rec, below=of FT]
  (final) {\begin{tabular}{c} Studies included for quantitative and qualitative analysis \\ \textbf{(n = 13)} \end{tabular}};

%5+6 label: Included

%\begin{comment}
\path let
\p1=(final.south), \p2=(Elli)
in
node[barstyle,left=of Elli,minimum width=\y2-\y1] 
  (barIncl) {Screening};
%\end{comment}
\begin{comment}
\path let
\p1=(final.south), \p2=(Elli.south -3em)
in
node[barstyle,left=of Elli.south west,minimum width=\y2-\y1] 
  (bar) {Included};
\end{comment}

%% Pfeile
\path[->,draw=black,line width=2pt]
    (DB) edge (dup)
    (dup) edge (screen)
    (screen) edge (ex)
    (screen) edge (FTin)
    (FTin) edge (FTout)
    (FTin) edge (Elli)
    (Elli) edge (Elliext)
    (Elli) edge (final);

\end{tikzpicture} % The file containing your standalone graph
    }
    \caption{Flowchart of the systematic literature review process: identification, screening, eligibility, and inclusion of articles. A white box implies inclusion and a grey box exclusion.}
    \label{fig:prisma_diagram}
\end{figure}

% [width=0.46\textwidth]

\subsection{Step 2 - Selection process}\label{sec:selection_process} 
The initial stage of the selection process serves to identify which literature should be included or excluded. Each of the 2,232 records is subjected to an individual screening based on its title and abstract. Following this, the literature is classified as either "inclusion" or "exclusion." Papers marked for "inclusion" advance to the subsequent filtering phase, whereas those designated for "exclusion" are removed from the review. In the event of uncertainty, the paper is included and subjected to further examination in the subsequent stage. The criteria for inclusion and exclusion at this stage are as follows:

\textbf{Inclusion Criteria} (must satisfy all)
\begin{enumerate}
    \item \textbf{Study Focus:} Papers that specifically focus on blockchain-based identity solutions with significant parts of the solution implemented on-chain.
    \item \textbf{Application Scope:} Works discussing or developing solutions applicable to on-chain applications within contexts such as decentralized finance (DeFi) or where a significant portion of the operational logic is blockchain-based, such as through smart contracts.
    \item \textbf{Publication Years:} Studies published earliest in 2017 up until today.
    \item \textbf{Language:} Papers written in English.
\end{enumerate}

\textbf{Exclusion Criteria} (may not match one)
\begin{enumerate}
    \item \textbf{Beyond Scope:} Papers that do not address blockchain-based identity solutions.
    \item \textbf{Minimal Blockchain:} Papers that utilize the blockchain as a medium for realizing off-chain identities
    \item \textbf{Quality Issues / Short Papers:} Abstracts, posters, presentations without full papers, or detailed methodology and results.
    \item \textbf{Superficial Solutions:} Papers that lack explanations for how the proposed solution is implemented in a technological or architectural sense.
    \item \textbf{No on-chain:} Rigorous solutions without on-chain applicability
    \item \textbf{Language Limitations:} Studies not available in English.
\end{enumerate}

By applying these criteria to screen the titles and abstracts, a total of \textbf{98} papers are identified as relevant and marked for inclusion.

\subsection{Step 3 - Eligibility Check} 

The objective of the eligibility check is to identify high-quality and relevant articles. To this end, the \textbf{98} full texts of all advanced articles are obtained. Each article is then subjected to a comprehensive examination, which entails reading the title, abstract, conclusion, and, if necessary, additional sections or the complete article. The final decision regarding inclusion is made based on the criteria defined in the preceding step. The article will be flagged if a researcher is uncertain about excluding an article after an extended period. In a quality control step, all flagged articles and a randomly selected subset will be processed and categorized again to verify their eligibility (test-retest). After completing step 3, "Eligibility Check," 15 articles remain and are moved on to the final step. The results are documented in the spreadsheet.

\subsection{Step 4 - Screening}
\label{subsec:screening}
The screening process marks each remaining article's methodological processing and information extraction. A set of screening questions (SQ) ensures a comprehensive and systematic information retrieval. These screening questions are designed to address the initial research questions and make the included articles more comparable. The existing spreadsheet is extended with columns for the screening questions. Each publication is read multiple times during the screening process to extract information from the paper and populate the spreadsheet with responses to the SQs. In some cases, the screening may be conducted by various researchers.

\subsection{Screening Questions}
\textbf{Technical Aspects:}
\begin{enumerate}[before=\vspace{4pt}, after=\vspace{4pt}, itemsep=2pt, label=\textbf{SQ\arabic*}]
    \item \textbf{DLT Technology:} Which Distributed Ledger Technology is used?
    \item \textbf{Additional Concepts:} Which supplementary concepts, standards, protocols, tools, or technologies are integrated, and for which components are these utilized?
    \item \textbf{Main Entities:} What main entities does the solution consist of?
    \item \textbf{On-chain versus Off-chain:} What components and entities are on-chain and off-chain?
    \item \textbf{Solution Maturity:} Is the solution theoretical, conceptually implemented, or deployed in a production environment?
    \item \textbf{Data Bridging Challenge:} What strategies are employed to address the challenge of making on-chain attestations based on real-world user data?
\end{enumerate} 

During the screening process, two additional articles were marked for exclusion, which leaves \textbf{13} articles relevant for the review.

\subsection{Outcome} 

\begin{figure}[!ht]
    \centering
    \begin{tikzpicture}
  \begin{axis}[
    xmax=2023.5,
    ymax=14,
    every axis plot/.append style={ultra thick},
    /pgf/number format/1000 sep={},
    ylabel=Number of Pulications,
    xlabel=Year,
    axis lines=middle,
    x label style={at={(axis description cs:0.5,-0.1)},anchor=north},
    y label style={at={(axis description cs:-0.1,.5)},rotate=90,anchor=south},
    axis lines=box,
    xtick={2017,2018,2019,2020,2021,2022,2023},
    ytick={1,3,5,7,9,11,13},
    xtick pos=left,
    ytick pos=left,
    ymin=0, xmin=2016.5, % Set the minimum values for the axes
    legend style={at={(0.5,-0.2)},anchor=north,legend columns=-1},
    cycle list name=color list, % Use the default cycle list
    ]
    \addplot [color=darkgray,mark=*] coordinates{
      (2017,1) (2018,3) (2019,3) (2020,3) (2021,6) (2022,9) (2023,13)
    };
    \addlegendentry{Cumulative of Publications}
  \end{axis}
\end{tikzpicture} % The file containing your standalone graph
    \caption{Cumulative number of publications on on-chain verifiable identities over the years, showing increasing research interest.}
    \label{fig:cumulative-publications-per-year}
\end{figure}
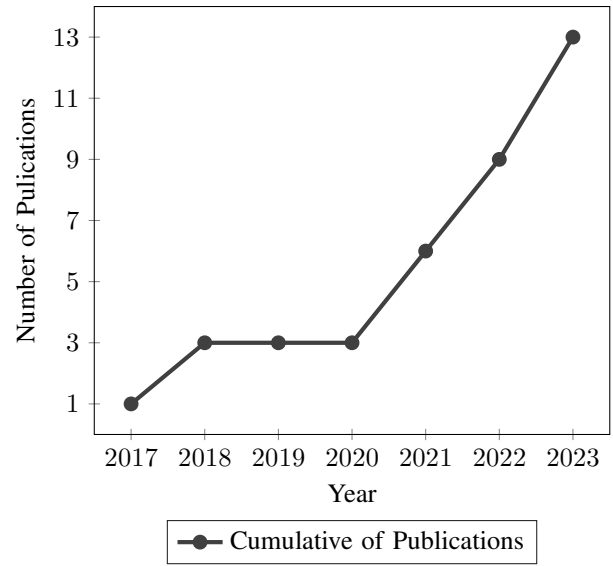

After conducting all the steps described above, 13 articles remained. These articles match the initially defined scope of this structured literature review and have been further processed to answer the initial research questions. During the filtration process, several observations could be made. The broad topic of identity and blockchain is an active research area with more than 2,232 publications matching the search term, as well as 98 articles being of general relevance to our narrowed-down criteria. However, only sparse literature exists in our specific subfield of on-chain verifiable identities. Figure \ref{fig:cumulative-publications-per-year} illustrates the growth in research interest in on-chain verifiable identities over time.

\section{Results}
\label{sec:results}
The results of the systematic literature review will be presented stepwise, with each previously introduced screening question addressed successively. The review data will be presented, and a qualitative assessment of the respective review findings will be performed.

\paragraph{SQ1 DLT Technology} \textit{Which Distributed Ledger Technology is used?} \\
Table~\ref{table:SQ1-DLT-Tech} provides an overview of the technology selected by different researchers. Some researchers combine two or more distributed ledger technologies to realize their systems or implement their solutions on different blockchain platforms. This results in some articles being counted more than once. As an example, \textit{Tadjik}~\etal~\cite{tadjik_blockchain_2022} uses a combination of Hyperledger Fabric and Hyperledger Indy in their protocol. Another example is \textit{Eres}~\etal~\cite{kai-jun-eres_bottom-up_2023} who utilize a combination of a public Ethereum network (Kovan testnet) and an additional permissioned Ethereum network (Hyperledger Besu). This article will be counted only once for "Ethereum", as both technologies are Ethereum networks.

\begin{table}[htp]
\caption{A summary of the distributed ledger technologies utilized or proposed in the examined research articles.}
\label{table:SQ1-DLT-Tech}
    \begin{center}
    %\begin{tabular}{M{1.7cm}|M{1.2cm}|M{4cm}}
    \begin{tabular}{p{2.9cm} p{3.4cm}} 
        \toprule
        \bfseries Technology &
        %\bfseries Articles & 
        \bfseries References  \\
        \midrule
        Ethereum \newline (\emph{9 articles}) & \cite{kai-jun-eres_bottom-up_2023}, \cite{azouvi_who_2017}, \cite{tavares_wallid_2018}, \cite{gallersdorfer_2021}, \cite{bruschi_privacy_2021}, \cite{heiss_non-disclosing_2022}, \cite{muth_towards_2023}, \cite{di_francesco_maesa_self_2023}, \cite{luong_privacy-preserving_2023} 
        \\ 
        \midrule
        Hyperledger Indy \newline (\emph{2 articles}) & 
        \cite{tadjik_blockchain_2022}, \cite{muth_towards_2023}
        \\
        \midrule
        Hyperledger Fabric \newline (\emph{2 articles}) & \cite{tadjik_blockchain_2022}, \cite{mukta_credtrust_2022} 
        \\
        \midrule
        Bitcoin \newline (\emph{2 articles}) &  
        \cite{azouvi_who_2017}, \cite{Buccafurri_integrating_2018} 
        \\
        \midrule
        Not specified \newline (\emph{1 articles}) &
        \cite{damgard_balancing_2021}
        \\
        \bottomrule
    \end{tabular}
    \end{center}

\end{table}

Hyperledger Indy and Fabric are used two times, respectively. Two solutions utilize Bitcoin. It should be noted, however, that these two solutions are two of the earliest publications included in our review, having been released in 2017 and 2018. One proposed solution did not specify the type of blockchain technology employed.

\paragraph{SQ2 Additional Concepts} \textit{Which supplementary concepts, standards, protocols, tools, or technologies are integrated, and for which components are these utilized?} \\ 
This screening question aims to examine the various approaches researchers employ in realizing on-chain identities. This analysis should provide insights and guidance for future researchers seeking to build upon these concepts. 

\begin{table}[htp]
\caption{Additional protocols, standards, tools, and technologies integrated into on-chain identity solutions.}
\label{table:SQ2-Additional-concepts}
    \begin{center}
    %\begin{tabular}{M{1.7cm}|M{1.2cm}|M{4cm}}
    \begin{tabular}{p{3.8cm} p{2.4cm}} 
        \toprule
        \bfseries Technology &
        %\bfseries Articles & 
        \bfseries References  \\
        \midrule
        zk-proofs \newline (\emph{7 articles}) & 
        \cite{kai-jun-eres_bottom-up_2023}, \cite{bruschi_privacy_2021}, \cite{heiss_non-disclosing_2022}, \cite{muth_towards_2023}, \cite{di_francesco_maesa_self_2023}, \cite{luong_privacy-preserving_2023}, \cite{damgard_balancing_2021}
        \\ 
        \midrule
        PKI / X.509 certificates \newline (\emph{3 articles}) & 
        \cite{azouvi_who_2017}, \cite{tavares_wallid_2018}, \cite{gallersdorfer_2021}
        \\
        \midrule
        WoT / Trust propagation \newline (\emph{3 articles}) & \cite{kai-jun-eres_bottom-up_2023}, \cite{azouvi_who_2017}, \cite{mukta_credtrust_2022}
        \\
        \midrule
        Other \newline (\emph{4 articles}) &  
        \cite{bruschi_privacy_2021}, \cite{muth_towards_2023}, \cite{luong_privacy-preserving_2023},  
        \cite{damgard_balancing_2021}
        \\
        \bottomrule
    \end{tabular}
    \end{center}
\end{table}

The most frequently utilized technical concepts are zero-knowledge proofs (zkps) in the form of zk-SNARKs, which have been employed in seven published works. Four of the six aforementioned publications, \cite{bruschi_privacy_2021}, \cite{heiss_non-disclosing_2022}, \cite{di_francesco_maesa_self_2023}, \cite{luong_privacy-preserving_2023} make use of the ZoKrates library for proof generation. Three solutions employ PKI and/or X.509 certificates. A comparison of the two approaches by publication date reveals that PKI/X.509 papers were published between 2017 and 2021, while zkp publications were released between 2021 and 2023. Three proposed solutions employed trust propagation or web of trust-related concepts. Additionally, four papers explicitly referenced the use of additional cryptographic protocols. Namely, \textit{Shamirs Secret Sharing} by \textit{Luong} \etal \cite{luong_privacy-preserving_2023} and \textit{Bruschi} \etal \cite{bruschi_privacy_2021}. The Camenisch-Lysyanskaya signatures are used by \textit{Damgard} \etal \cite{damgard_balancing_2021} and \textit{Muth} \etal \cite{muth_towards_2023}. In addition, the Pointcheval-Sanders signature scheme and Pedersen commitments are employed by \textit{Damgard} \etal \cite{damgard_balancing_2021} and MiMc block ciphers are utilized by \textit{Bruschi} \etal \cite{bruschi_privacy_2021}.

\paragraph{SQ3 Main Entities} \textit{What main Entities does the solution consist of?} \\ 
The majority of the authors refer to an issuer-holder-verifier model. All mention a "user," "holder," or "account" or a similar form of the subject of the credentials, who holds and controls them. However, depending on the proposed framework, there are some modifications to this model. Similarly, every examined work mentions an issuer of a credential. The issuer acts as a trusted entity that verifies the identity of the subject and creates the digital credentials. Depending on the context, the issuer is referred to as a registrar, a trusted entity, an identity provider, an attribute manager, or sometimes a trust anchor. Furthermore, all proposed solutions involve some verifier or verification. Nevertheless, not all explicitly mention them by name or utilize new entity names, often called service providers, to act as verifiers.
Other entities of interest are higher-level instances with extended rights within the system. Six frameworks \cite{gallersdorfer_2021, mukta_credtrust_2022, bruschi_privacy_2021, damgard_balancing_2021, kai-jun-eres_bottom-up_2023, tavares_wallid_2018} rely on the concept of \textbf{"trust anchors"}. Also referred to as a "registry", "governance authority", "maintainer", "trusted third party", or "credible entity". Other components mentioned are a "Judicial Authority" \cite{bruschi_privacy_2021} and "Anonymity Revoker" \cite{damgard_balancing_2021}. These privileged entities can reveal a user's real-world identity in some cases of fraudulent activity.

\paragraph{SQ4 On-chain versus Off-chain} \textit{What components and entities are on-chain and off-chain?} \\ 
%off-chain verification solutions
Five out of thirteen authors implement off-chain-centric identity verification. \textit{Mukta}~\etal~\cite{mukta_credtrust_2022} and \textit{Azouvi}~\etal~\cite{azouvi_who_2017} WoT based solutions with on-chain trust registries. In both solutions, trust evaluation and verification are handled by off-chain logic. \textit{Buccafurri}~\etal~\cite{Buccafurri_integrating_2018} only superficially mentions the verification process, which is done by querying past registration transactions and validating the corresponding cryptographic keys off-chain. \textit{Tavares}~\etal~\cite{tavares_wallid_2018} uses a certificate-based approach, where certificates are stored on-chain but validated off-chain. \textit{Tadjik}~\etal~\cite{tadjik_blockchain_2022} uses Hyperledger Indy, which is a blockchain designed for the sole purpose of holding identities. However, the resulting data must be accessed and utilized off-chain. Consequently, this approach is considered off-chain-centric verification.

%on-chain verification solutions
Eight solutions implement partial or full on-chain verification of identities.
%Bruschi %Luong
\textit{Heiss} \cite{heiss_non-disclosing_2022},  \textit{Luong} \cite{luong_privacy-preserving_2023}, \textit{Bruschi} \cite{bruschi_privacy_2021}, \textit{Di Francesco Maesa} \cite{di_francesco_maesa_self_2023}, \textit{Muth}~\etal~\cite{muth_towards_2023} and their respective colleagues propose a zk-SNARK-based on-chain verification method. Their approaches involve calculating an attributed/identity proof off-chain and submitting it to a smart contract for on-chain verification. Upon successful proof verification, the user may be added to a list of verified entities or marked as verified.
%Jun
\textit{Jun}~\etal~\cite{kai-jun-eres_bottom-up_2023} employ a combination of a public and a private ledger. The trust graph of their WoT-based approach is stored on-chain in their private, access-restricted blockchain. Trust calculations, identity verification, and most other calculations are conducted off-chain. However, a non-interactive zkp is published on the public ledger, allowing on-chain credential verification.
%Damgard
The protocol by \textit{Damgard}~\etal~\cite{damgard_balancing_2021} issues each account holder with an off-chain certificate, which allows for creating new accounts. This certificate is obtained from an identity provider. Publicly unlinkable accounts are generated by using this certificate. The anonymity of accounts can be revoked by collaborating with a specific number of anonymity-revokers off-chain. The creation information is submitted on-chain after the off-chain generation of a new account. This information contains a zkp verifying the correctness of the account and confirming that an identity provider signed the attributes published by the user. The number of attributes the user wishes to include is at their discretion.
%Gallersdörfer
\textit{Gallersdorfer}~\etal~\cite{gallersdorfer_2021} employs a certificate-based approach, wherein X.509 certificates are stored and verified on the blockchain. This encompasses the processes of certificate creation, modification, retrieval, and revocation and the interrelationships between certificate chains.

%Todo: folgende question "life system" anpassen
\paragraph{SQ5 Solution Maturity} \textit{Is the solution theoretical, conceptually implemented, or deployed in a production environment?} \\ 
All proposed systems except one were designed and deployed as a conceptual implementation (proof-of-concept). Only \textit{Damgard}~\etal~\cite{damgard_balancing_2021} proposed a theoretical system without an actual implementation. Some researcher tested their deployment time. However, deployment time varies depending on the actual blockchain usage. Therefore, deployment time is less relevant in the public blockchain context since the most significant variable of public ledgers is gas costs. In addition, most systems do not face a time constraint but rather a cost constraint. For example, the solutions of \cite{bruschi_privacy_2021} and \cite{muth_towards_2023} consume several million gas units for on-chain proof verification. The solution of \textit{Heiss}~\etal~\cite{heiss_non-disclosing_2022} consumes about 600k\footnote{About 30 USD with a gasprice of 15 GWei and an Ethereum price of 3500 USD} gas for proof verification, which is achieved by off-chain pre-processing of the identity. However, the deployment costs of the verifier smart contract are not mentioned. \textit{Gallersdorfer}~\etal~\cite{gallersdorfer_2021} consumes 500k - 1.5M units of gas for on-chain PKI certificate submission.

\paragraph{SQ6 Data Integration Challenge} \textit{What strategies are employed to address the challenge of making on-chain attestations based on real-world user data?} \\
The implementation of on-chain identity for real-world actors necessarily involves using some real-world data for on-chain attestations (e.g. ensuring an identity) based on this data. Consequently, there is a need for trust in the data. This raises the question of how identities are verified and how the root trust of blockchain actors is guaranteed. Three distinct approaches could be observed: \one protocol does not address this question at all \cite{tavares_wallid_2018, heiss_non-disclosing_2022}. \two introduce some trust anchor or trusted authority without further mentioning the implementation \cite{tadjik_blockchain_2022, luong_privacy-preserving_2023}. \three using a governmental digital id \cite{tavares_wallid_2018}. Nevertheless, the question of how the trust of identity documents and on-chain entities is established within those protocols remains unresolved.
\section{Discussion}
\label{sec:discussion}

% Summary: A brief recap of your key results
In the preceding section \ref{sec:results}, the results of the analysis of the thirteen included articles were presented by answering the six previously defined screening questions. 
%Q1
It can be seen that a majority of solutions are implemented using Ethereum-based ledgers. 
%%Q4
The initial identity verification of a new user and most cryptographic computations, such as key exchange, hashing, proof generation, certificate creation, and similar setup steps, are conducted off-chain. The most widely used on-chain components involve some level of identity or credential registration. Entities participating in the system may also be registered on-chain. Eight solutions allow on-chain, smart contract-based verification of an identity claim [\ref{RQ1}].
%%Q5
All but one implemented a proof of concept for their solution, and various researchers tested their deployment.
%% Q2
The most prominent technologies for realizing some form of on-chain identity verification are non-interactive zero-knowledge proofs, which were employed by seven researchers, as well as PKI or WoT solutions, which were present in three architectures, respectively [\ref{RQ2}].
%% Q3
Most solutions adhere to a model in which the issuer, holder, and verifier are the primary actors within the system. 
%Q6
Research on incorporating initial trust into the system is still awaiting a thorough investigation by a researcher. This lack of trust manifests itself in two different ways. First, it is necessary to determine how the issuer of an on-chain identity can be trusted. Second, it is essential to investigate how the user can verify their "real-world identity" to the issuer [\ref{RQ3}].
% Interpretations: What do your results mean?

Ethereum was chosen for 70\% of all solutions, possibly due to its established smart contract functionality, comprehensive developer tools, and mainstream adoption. However, due to the high cost of on-chain computation and storage, all authors outsource most of their protocol off-chain. 
The most notable distinction between the solutions under consideration can be observed in the on-chain credential verification phase, where zkps, WoT, or PKIs are employed on-chain. However, no solution would be suitable for a live Ethereum mainnet implementation. The reasons for this are as follows: \one the high gas costs associated with zkps, \two unrealistic trust assumptions of WoT or PKI, where the question of initial trust in the system is unsatisfactorily answered, and \three the missing consideration of how real-world identities are bridged into the blockchain.

% Implications: Why do your results matter?
% Recommendations: Avenues for further studies or analyses
The research area of blockchain combined with identity management is a topic of interest, with 98 articles generally matching the specific topic. However, only 13 articles were suitable for this research, indicating a niche with increasing publications over the last years. A comprehensive overview of the state of the art, technologies used, and potential future developments in on-chain identity solutions was presented. For future research, several distinct research directions can be proposed. First, zkps proof sizes and computations are currently infeasible for on-chain applications. Therefore, further research is needed to make zkps more efficient at the algorithmic level. An alternative approach would be to enhance the integration of zkps within the blockchain protocol. Secondly, the trust issue represents a significant challenge that requires further investigation. The lack of trusted mechanisms on public blockchains has resulted in the absence of trusted anchors that can verify new entities. Third, the next trust issue is the privacy-compliant bridging of real-world human identities into the digital space and possibly on-chain. Most researchers have not explicitly explained how trust or identities are transferred to the blockchain. Therefore, we recommend additional research focusing on the certification of blockchain actors, on-chain trusted entities, or other novel approaches to introduce trust anchors on permissionless distributed ledgers.
\section{Conclusion}
\label{sec:conclusion}

This systematic literature review investigated the establishment of blockchain-based (on-chain) identities and synthesized existing research to identify prevailing trends, gaps, and implications. On-chain identity was defined as the data protection-compliant on-chain attestation of a verified identity without storing personal data on the blockchain. Articles were included in the review when a significant part of the logic was implemented on the blockchain or when identities could be verified within the logic on-chain. The systematic review revealed that most research articles covering the areas of blockchain and identity did not meet the defined criteria of on-chain identity. This was largely due to the fact that the blockchain was merely used as a secure storage for cryptographic hashes for similar data. 

By conducting this literature review, a rigorous and structured approach was employed, guided by the PRISMA framework, to limit subjectivity and bias as much as possible. Quality checks were incorporated to ensure the integrity of the data. It should be noted that this study has some limitations. Firstly, subjectivity was introduced during the formation of the search term, the definition of the inclusion criteria, and the decision of which studies to include. Secondly, the literature included is insufficient for quantitative claims but well-suited for qualitative analysis.

The majority of included articles follow an issuer-holder-verifier model and use Ethereum or Hyperledger Fabric / Indy for their deployment. Further utilized technologies are zkps, PKI / X.509 certificates, and WoT approaches. A notable absence in the literature is an examination of the mechanisms by which trust is established within the context of on-chain actors. Another significant gap in the literature is the identification and digitization of human identity data, which must be trusted and compliant.

Future research should investigate methods to enhance the efficacy of zero-knowledge proofs. Potential avenues for exploration include enhancing the proof algorithms, extending the blockchain protocol with dedicated zkp functionality, and introducing new interoperability architectures of zkps and blockchains. Additionally, there is a need to address the lack of consideration of trust. This highlights two additional research areas: first, the identification and verification of real-world identities before issuing digital representations, and second, the establishment of trust in on-chain actors issuing blockchain-based identities.

\ifCLASSOPTIONcaptionsoff
  \newpage
\fi

\newpage
\bibliographystyle{./bibtex/IEEEtran}
\bibliography{./bibtex/bib/IEEEabrv,./bibtex/bib/IEEEreferences,./bibtex/bib/IEEEreference_13included,bibtex/bib/IEEEreferences_background}

%\vfill


\end{document}